\documentclass[12pt,preprint]{aastex}
\usepackage{lscape}
\usepackage{multirow}

\def\lsim {$\rlap{\raise.4ex\hbox{$<$}}\lower.55ex\hbox{$\sim$}\,$}

\newcommand{\neff}{n$_{\mathrm{eff}}$}

\newcommand{\hcop}{HCO$^+$}

\newcommand{\hcopi}{H$^{13}$CO$^+$}

\begin{document}

               
\title {\bf Evidence for Inflow in High-Mass Star-Forming Clumps}
\author {Megan Reiter, Yancy L. Shirley\altaffilmark{1}}
\affil{Astronomy Department, The University of Arizona,
       933 N. Cherry Ave., Tucson, AZ 85721 \\
mreiter@as.arizona.edu, yshirley@as.arizona.edu}

\author {Jingwen Wu}
\affil{Jet Propulsion Laboratory, 
       169-306, 4800 Oakgrove Drive, Pasadena CA 91109 \\ 
jingwen.wu@jpl.nasa.gov}

\author{Crystal Brogan, Alwyn Wootten}
\affil{National Radio Astronomy Observatory, 520 Edgemont Road,
       Charlottesville, VA 22903 \\
cbrogan@nrao.edu, awootten@nrao.edu}
\and
\author{Ken'ichi Tatematsu}
\affil{National Astronomical Observatory of Japan, 2-21-1 Osawa, Mitaka
       Tokyo 181-8588, Japan \\
k.tatematsu@nao.ac.jp}

\altaffiltext{1}{Adjunct Astronomer at the National Radio Astronomy 
Observatory. The National Radio Astronomy Observatory is a facility of the 
National Science Foundation operated under a cooperative agreement by 
Associated Universities, Inc.}
 
\begin{abstract}
We analyze the \hcop\ 3-2 and \hcopi\ 3-2 line profiles of 27 high-mass star-forming regions to identify asymmetries that are suggestive of mass inflow. 
Three quantitative measures of line asymmetry are used to indicate whether a line profile is blue, red or neither --- 
the ratio of the temperature of the blue and red peaks, the line skew and the dimensionless parameter $\delta v$. 
We find nine \hcop\ 3-2 line profiles with a significant blue asymmetry and four with significant red asymmetric profiles. Comparing our \hcop\ 3-2 results to HCN 3-2 observations from Wu et al. (2003, 2010), we find that eight of the blue and three of red have profiles with the same asymmetry in HCN. 
The eight sources with blue asymmetries in both tracers are considered strong candidates for inflow. 
Quantitative measures of the asymmetry (e.g. $\delta v$) tend to be larger for HCN. 
This, combined with possible \hcop\ abundance enhancements in outflows, suggests that HCN may be a better tracer of inflow. 
Understanding the behavior of common molecular tracers like \hcop\ in clumps of different masses is important for properly analyzing the line profiles seen in a sample of sources representing a broad range of clump masses. Such studies will soon be possible with the large number of sources with possible self-absorption seen in spectroscopic follow-up observations of clumps identified in the Bolocam Galactic Plane Survey. 
\end{abstract}



\section{Introduction}

Isolated, low-mass star formation is generally accepted to result from the gravitational collapse of a dense core within a molecular cloud (e.g. Shu, Adams \& Lizano 1987). In this picture, cores grow by inside-out collapse with material inflowing with higher velocities near the core. 
Evidence of collapse is observable in self-absorbed, optically thick line profiles. 
Emission from the infalling envelope on the far side of the protostar will be blueshifted by an amount proportional to the velocity gradient towards the core. This blueshifted emission will not be absorbed by foreground layers that are warmer or at a substantially different velocity (see Walker et al. 1986; Walker, Narayanan \& Boss 1994), leading to excess emission in the line profile blueward of the source velocity. 
For moderately optically thick sources, the line will appear skewed blueward while strongly self-absorbed sources will show two distinct peaks with the blue peak moderately to significantly brighter than the red. 
Large temperature gradients in the core will increase the depth of the self-absorption feature while large velocity gradients will increase the asymmetry of the line. 

Before a source with a blue asymmetric line profile can be considered a collapse candidate, other mechanisms for creating a blue profile must be ruled out. 
Spatially coincident cores that are not physically associated, rotation and outflows can all produce line asymmetries. 
Emission from an optically thin isotopologue will peak in the middle of the absorption dip of the optically thick line if the emission is from a single source. 
Rotation and outflows also generate double-peaked line profiles, though their contribution to the line profile is more apparent with maps of sufficient spatial resolution. Even in the absence of such maps, both rotation and outflows should produce equal numbers of red and blue asymmetric line profiles. 
Because line profiles may simultaneously include contributions from outflows, expansion and rotation in addition to inflow, there are conflicting views on what constitutes unambiguous evidence for inflow (i.e. Snell \& Loren 1977; Leung \& Brown 1977). 
High resolution observations have spatially resolved inflowing, outflowing and rotating components for a small number of sources (e.g. Remijan \& Hollis 2006, Keto \& Wood 2006). 
Other sources are considered strong candidates for infall based on observations of blue asymmetries in multiple molecular lines, probing a range of densities. 
For a sample of sources, a statistically significant excess of blue profiles suggests that inflow may be an important physical process in determining the line profile.

Previous work on infall has either focused on detailed study of a single source (e.g. IRAS 16293-2422, Walker et al. 1986; Menten et al. 1987; Walker et al. 1988; Narayanan, Walker \& Buckley 1998; Ceccarelli et al. 2000; Chandler et al. 2005; Remijan \& Hollis 2006) or statistical studies of a large sample of sources (e.g. Mardones 2003). 
While multiple statistical studies have examined candidates for infall among low-mass stars (e.g. Mardones et al. 1997; Gregersen et al. 1997; Gregersen et al. 2000; Lee \& Myers 2011), the literature on higher mass candidates is less extensive and typically only one kinematic tracer is analyzed (e.g. Wu \& Evans 2003; Fuller et al. 2005; Klaassen \& Wilson 2008; Sun \& Gao 2009; Cyganowski et al. 2009). 
Massive stars form in complex environments and at higher median distances than low mass sources, thus features of individual cores embedded in the larger clump are averaged together in the single-dish beam, making infall motions harder to observe. 
Observations of high-mass clumps are also complicated by strong temperature gradients and outflows from the multiple cores likely embedded in the clump.
However, gas dynamics of the clump envelope may help discriminate between proposed modes of high mass star formation. For example, cores of moderate mass may grow to become massive stars via the global inflow of clump material because of their location deep in to the cluster potential (Smith, Longmore \& Bonnell 2009).

In this work, we analyze the line profiles of 27 high-mass star-forming regions observed in \hcop\ and \hcopi\ 3-2 for evidence of inflow. 
New \hcopi\ 3-2 observations are described in \S2. 
We examine these line profiles for asymmetry and calculate the excess of blue profiles relative to red profiles in \S3. 
Noting that every source with a blue asymmetry in \hcop\ 3-2 is also blue asymmetric in HCN 3-2 (Wu \& Evans 2003; Wu et al. 2010), we compare our \hcop\ results to those found for the same sources in HCN 3-2 in \S4. We discuss these conclusions and their implications for the advent of ALMA in \S5.

\section{Observations}
We use the 
\hcop\ 3-2 ($\nu = 267.5576190$ GHz) and \hcopi\ 3-2 ($\nu = 260.2553390$ GHz) observations published in Reiter et al. (2011). 
We compare these \hcop\ line profiles to the HCN 3-2 ($\nu = 265.886188$ GHz) and H$^{13}$CN 3-2 ($\nu = 259.011814$ GHz) profiles presented by Wu et al. (2003, 2010). 

We also observed the 7 sources not detected in \hcopi\ 3-2 in Reiter et al. (2011) with the 1mm ALMA prototype sideband-separating receiver on the Heinrich Hertz Telescope located on Mt. Graham, Arizona on February 15, 2011 (see Table 2). Observations were done in dual polarization 4 IF mode (6 GHz IF) using the filterbanks (1.102 km s$^{-1}$ resolution). Pointing and main beam efficiency calibrations were done with Venus ($\eta_{mb}^{V_{pol}}=0.71$ and $\eta_{mb}^{H_{pol}}=0.61$). 
Spectra were reduced using GILDAS CLASS reduction software. 
The calibration techniques used for HHT data are described in detail in Reiter et al. (2011).

\section{Results}\label{s:results}

\begin{figure}[h!]
\epsscale{0.9}
\includegraphics[angle=0,scale=0.60]{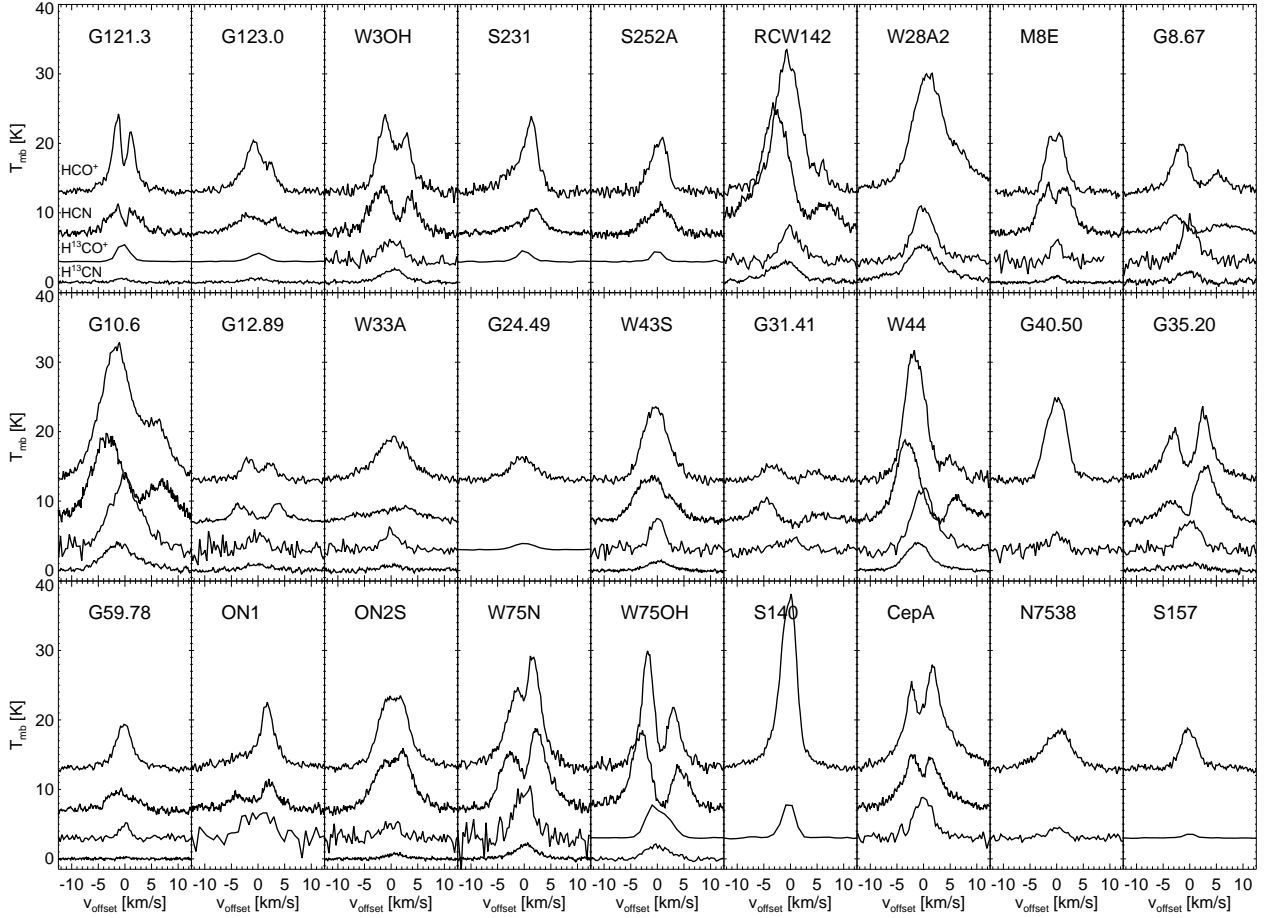}
\figcaption{Spectra from which we measure line asymmetries: in order from top to bottom, they are \hcop\ 3-2, HCN 3-2 (From Wu et al. 2010), \hcopi\ 3-2 (multiplied by two) and H$^{13}$CN 3-2 (also from Wu et al. 2010). An offset has been added to the \hcop, \hcopi\ and HCN spectra for clarity. 
Temperatures are plotted as a function of velocity offset from the v$_{LSR}$ determined from \hcopi\ 3-2 (see Table 2, Reiter et al. 2011). }\label{f:spectra}
\end{figure}

Table 1 lists the classification of the \hcop\ line profile for all of the sources in this sample. By visual inspection, we classify nine sources (33\%) as blue asymmetric and four as red asymmetric (15\%). 
The line asymmetry is strongest at the center positions for all 13 sources, although the asymmetry is still apparent a few arcseconds away for many of the sources (see Table 1, Figure 2). 
More quantitatively, the ratio of the temperature of the blue peak to the red peak, $T_R^*(B) / T_R^*(R)$, should be greater than one for blue profiles while it will be less than one for red profiles. 
By fitting a Gaussian to each peak independently, we find the red and blue peak temperatures (see Table 1, Figure 3).  
Eight of the blue profiles and all four of the red profiles have temperature ratios $\geq 1 \sigma$ and are considered significantly asymmetric.

\begin{figure}[h!]
\epsscale{0.9}
\includegraphics[angle=0,scale=0.45]{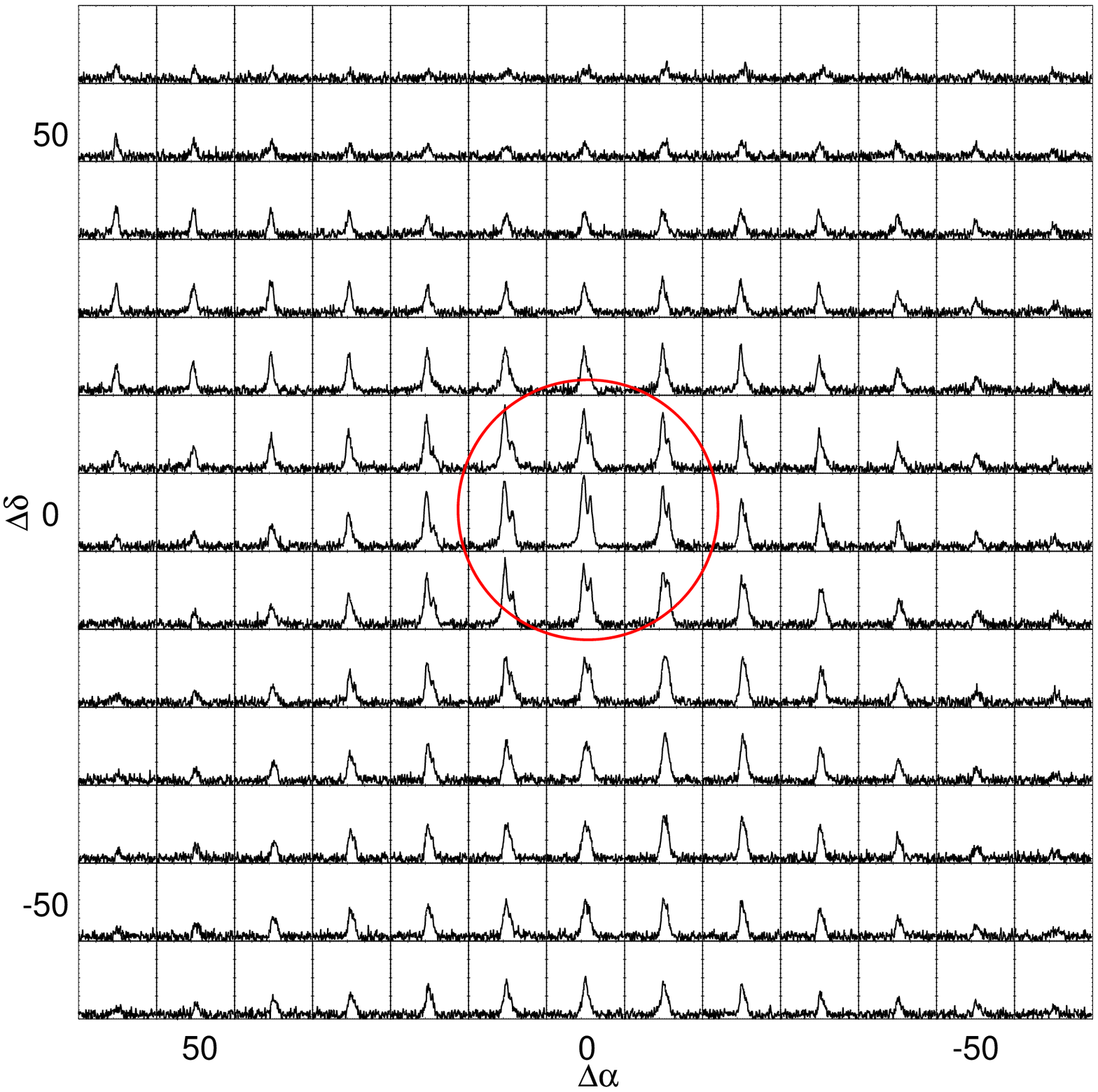}
\figcaption{A grid of all \hcop\ 3-2 spectra in the map of W3(OH) with a red circle indicating the beam size. See Table 1 for the median radius to which the line asymmetry persists. }\label{f:sample_spec}
\end{figure}

Other methods for quantifying the line asymmetry are useful for classifying line profiles that are clearly skewed but do not have two well-defined peaks. We calculate the skewness as 
\begin{equation}\label{e:skew}
skew = \frac{ \frac{\sum (T_A^* (v-v_{LSR})^3 \Delta v_{chan})}{\sum (T_A^* \Delta v_{chan})}} {\left( \frac{\sum(T_A^* (v-v_{LSR})^2 \Delta v_{chan})}{\sum (T_A^* \Delta v_{chan})}\right)^{3/2}}
\end{equation}
where $v_{LSR}$ is the velocity of the central source and has been determined from the isotopologue (which is presumed to be optically thin), and $\Delta v_{chan}$ is the channel width (Gregersen et al. 1997).
Lines with a blue asymmetry will have a negative skew.

Following Gregersen et al. (1997), we calculate the skewness over the velocity interval $v_{LSR} \pm 1.25$ km s$^{-1}$, which corresponds to a velocity width slightly larger than half of the median isotopologue linewidth ($\sim 2.2$ km s$^{-1}$ FWHM).
We require that the skewness be $\geq 3$ times its standard deviation to be considered significant. 
Eight of the nine visually identified blue profiles have a significant negative skew, in agreement with the visual classification, while of the two profiles with a significant red skew, only one was visually classified as a red profile. 

The velocity interval used by Gregersen et al. (1997) was determined from a sample of low-mass cores, all of which have smaller linewidths than our high-mass clumps. 
Using the same velocity range for our high-mass sample often misses one or both peaks of the clearly self-absorbed lines (see Figure 1). 
Expanding the velocity range to the full isotopologue linewidth, $v_{LSR} \pm \frac{1}{2} \Delta v_{iso}$ km s$^{-1}$, we find the same general trend in the skews as we found using the smaller velocity interval, however the uncertainty is now larger than the skew in every case (see Figure 3). 
This will be a general problem for all but the highest signal-to-noise spectra and is a fundamental limitation in the calculation of skewness.

Alternatively, we can calculate the dimensionless parameter 
\begin{equation}
\delta v = (v_{thick} - v_{LSR}) / \Delta v_{thin}
\end{equation}
where the difference between $v_{thick}$, the velocity of the brighter peak of the optically thick line, and $v_{LSR}$, determined from the optically thin isotopologue, is normalized by the linewidth of the optically thin isotopologue $\Delta v_{thin}$ (Mardones et al. 1997). Mardones et al. (1997) argue that $\delta v$ is less sensitive to asymmetries in the line wings which may be due to motions not associated infall (e.g. outflows). 
A significantly red or blue asymmetric line profile has $| \delta v | \geq 0.25$ which corresponds to a separation between the peak of the optically thick line and the $v_{LSR}$ of more than a quarter of the optically thin linewidth (Mardones et al. 1997).

Six of the nine blue asymmetric profiles identified with other methods meet the $| \delta v | \geq 0.25$ criteria for blue asymmetry (see Figure 4). 
Five sources meet the criteria for red asymmetry. Three of these were visually identified as red profiles. 
Skew and $\delta v$ agree for all sources where both measures of asymmetry are significant.

\begin{figure}[h!]
\epsscale{0.9}
$\begin{array}{cc}
\includegraphics[angle=90,scale=0.30]{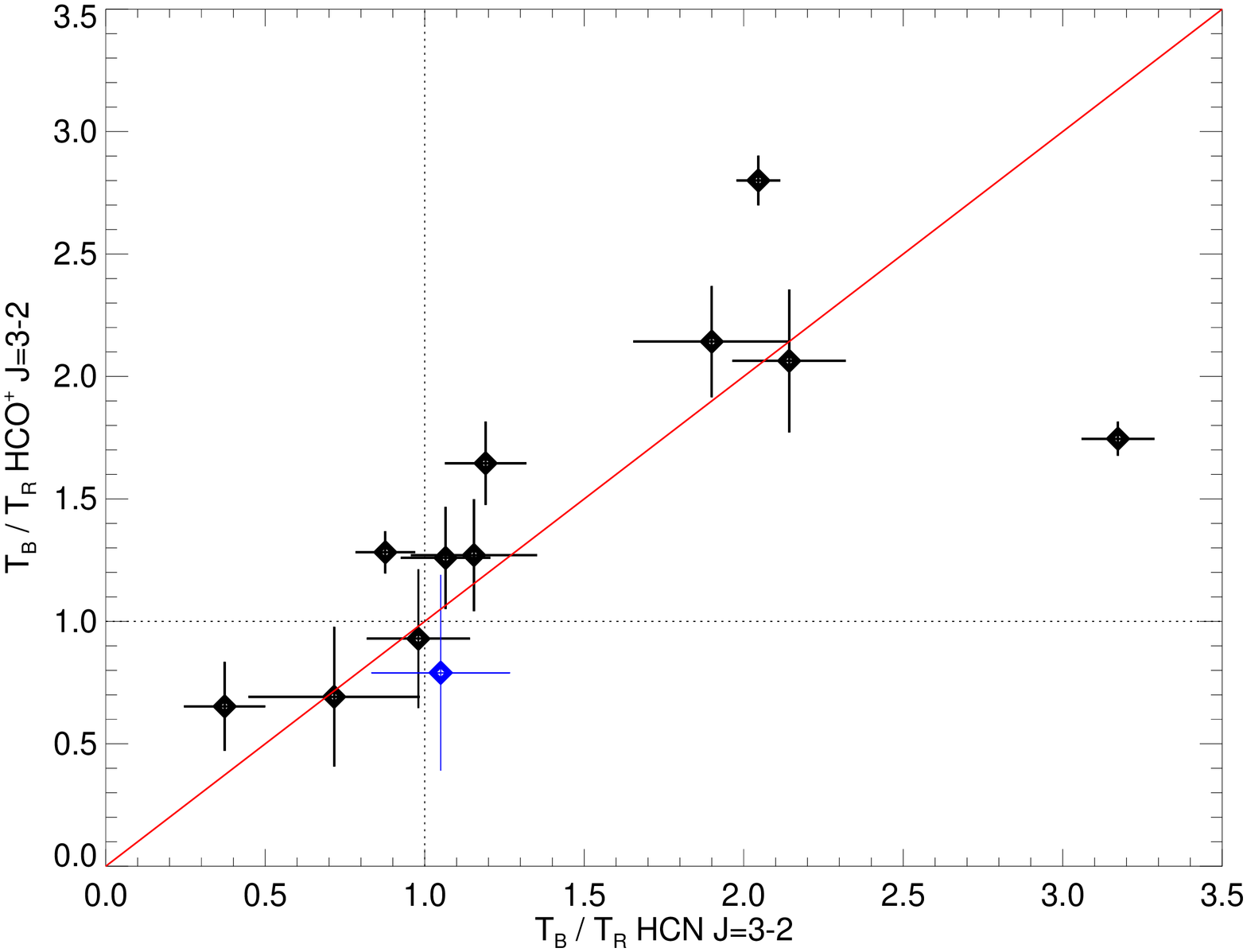} & 
\includegraphics[angle=90,scale=0.30]{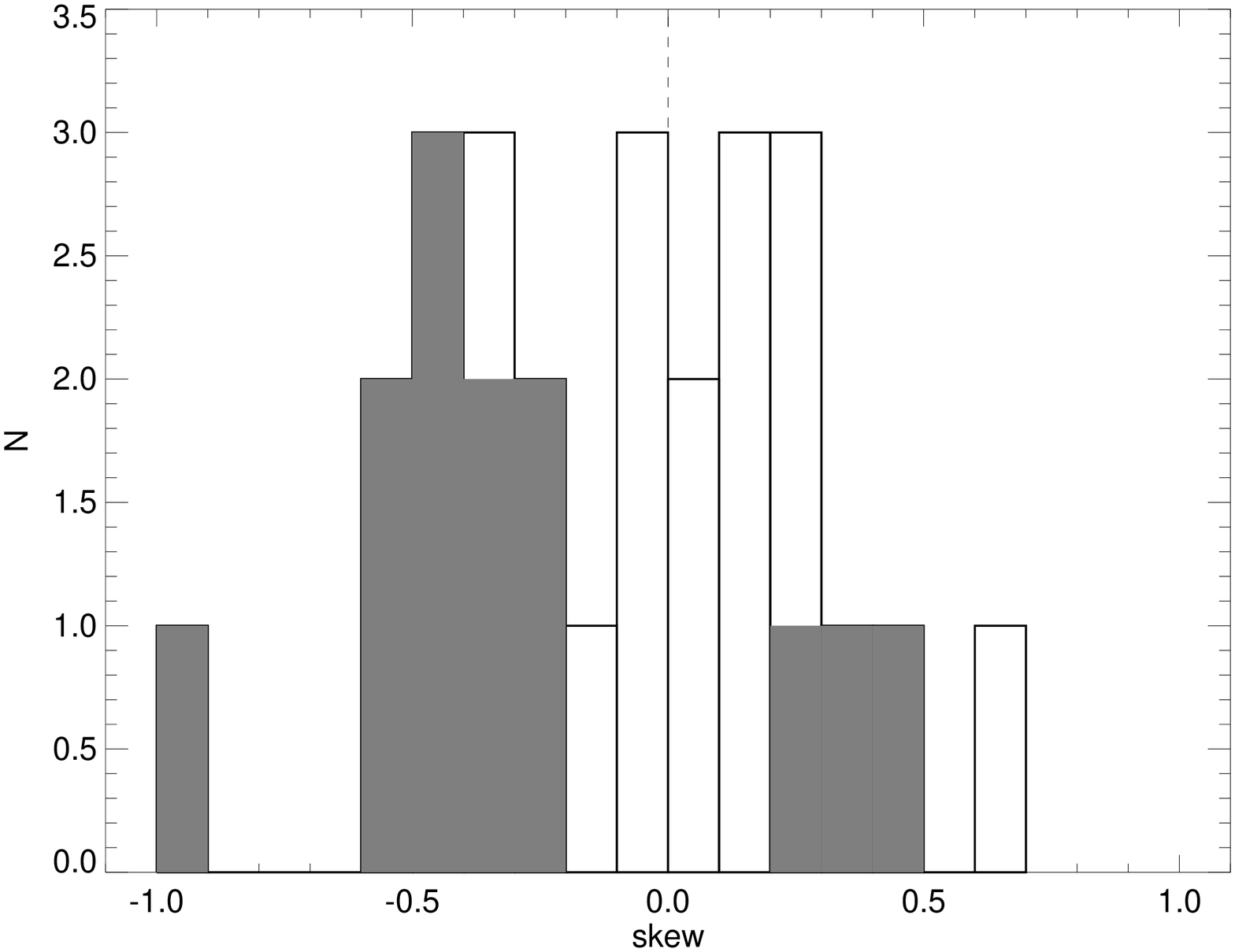} \\
\end{array}$
\figcaption{Left: Blue-to-red temperature ratio $T_R^*(B) / T_R^*(R)$ for \hcop\ and HCN. The blue point is CepA, the only source that changes from blue (HCN) to red (\hcop) asymmetry between tracers. 
Right: Histogram of the skew values for the \hcop\ profiles using a velocity interval $V_{LSR} \pm 1.25$ km s$^{-1}$. Solid lines indicate the histogram of all skews while the shaded bars represent the skews that meet the significance criterion (see \S~\ref{s:results}).}\label{f:tb_tr_n_skew}
\end{figure}

Deciphering the relative contribution of inflow, rotation and outflows to the observed line profiles is not possible at these resolutions (30\arcsec\ corresponding to $\sim 84,000$ AU at the median distance of 2.8 kpc). For a sample of sources, we expect that rotation and outflows should produce equal numbers of red and blue profiles, whereas inflow will only produce blue profiles. 
To quantify whether there is a statistical excess of blue profiles, we use the ``blue excess'' defined in Mardones et al. (1997) as 
\begin{equation}
E = (N_{blue} - N_{red}) / N_{total} 
\end{equation} 
where $N_{blue}$ and $N_{red}$ are the numbers of blue and red profiles while $N_{total}$ is the number of sources in our sample. 
The blue excess obtained using significantly asymmetric profiles identified with the three classification methods --- temperature ratio ($\geq 1\sigma$), skew ($\geq 3 \sigma$), $\delta v$ ($\geq 0.25$) --- yield $E=0.15, 0.30, 0.04$ respectively. 
Requiring that a profile is significantly asymmetric by two of the three methods leads to an overall blue excess of $E = 0.19$. 
These numbers (except for the $\delta v$ blue excess) are similar to those obtained by Wu \& Evans ($E=0.29$ for temperature ratio, $E=0.21$ for $\delta v$; Wu \& Evans 2003) in their preliminary analysis of the HCN 3-2 profiles of 28 high-mass star-forming regions, many of which are also observed in our sample (see \S~\ref{s:comparison}). 
%
%
Blue excess tends to be somewhat larger for low mass star forming regions observed in \hcop\ 3-2 ($E \approx 0.30$; Evans 2003), although Mardones et al. (1997) found a smaller blue excess for a sample of $47$ low mass star forming regions, observed in H$_2$CO and CS.
Not all surveys demonstrate a blue excess (e.g. Mardones 2003), although smaller blue excesses may be explained by differences in the optical depth between tracers (Evans 2003, Wu et al. 2010).
Sample selection may also bias the blue excess.
However, for a sample of high-mass star forming regions, Wu et al. (2010) find that asymmetries in HCN 3-2 are often not reflected in the CS 2-1 profiles. Those authors conclude that CS 2-1 is not sufficiently optically thick to trace the inner regions of the clump where inflow occurs.

A nearly equal number of red (5 of 27) and blue (6 of 27) asymmetries, as we find with the $\delta v$ parameter (see Table 1), is consistent with a uniform distribution where blue, red and symmetric profiles are all equally likely. 
A Monte Carlo simulation that picks 27 points from this uniform distribution will reproduce the $\delta v$ result for $\sim$90\% of 10000 trials. 
Even for the visual classification, which has a more pronounced excess of blue profiles (9 of 27) compared to red (4 of 27), the probability that our results could be obtained from a uniform distribution of line profiles is still $\sim$50\%. 
Quantitative measures of the asymmetry agree on the general shape of the line profile, but those results are not always significant, even for clearly asymmetric sources (e.g. G123.0 and W3(OH); see Figure 1, Table 1). 
Furthermore, line asymmetries do not always agree between different transitions of the same molecule (e.g. Gregersen et al. 1997). 
Additional observations in other molecules and at higher resolution are evidently necessary before a single source can be considered a strong candidate for collapse. 
In the next section, we compare our \hcop\ line asymmetry results to previous studies of line asymmetries in these sources using other molecular tracers.

\begin{figure}[h!]
\epsscale{0.6}
$\begin{array}{cc}
\includegraphics[angle=90,scale=0.30]{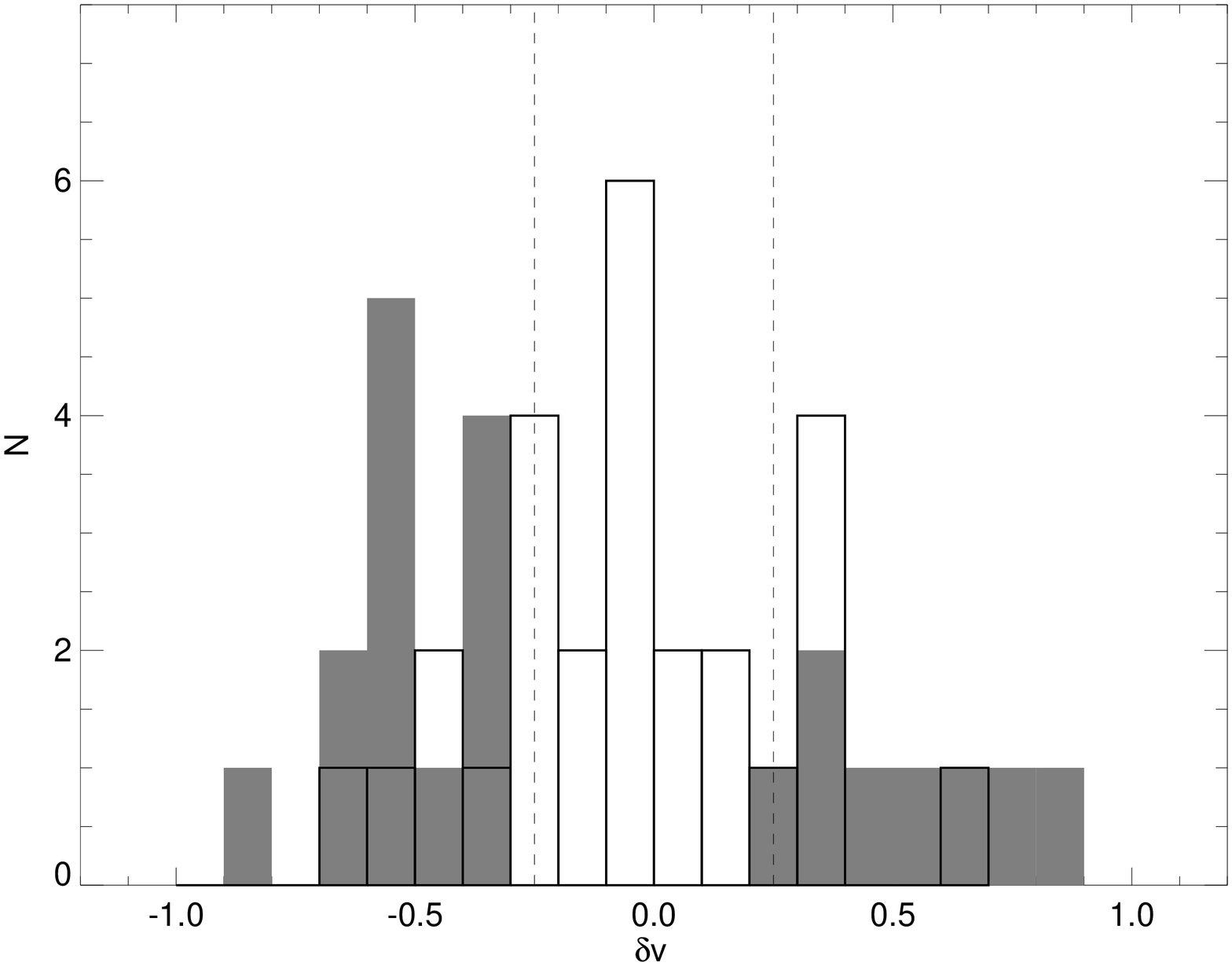} & 
\includegraphics[angle=90,scale=0.30]{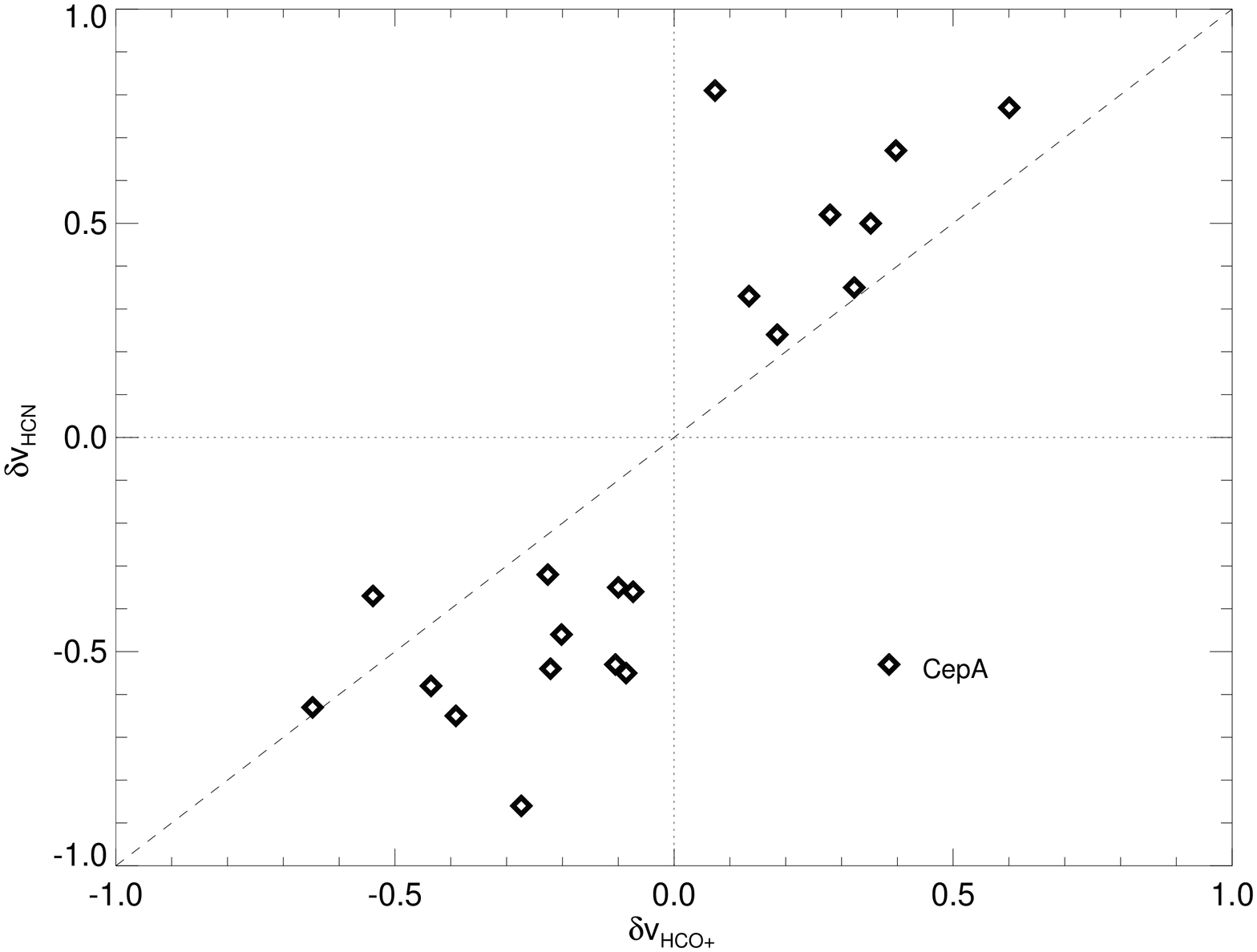} \\ 
\end{array}$
\figcaption{Left: Histogram of $\delta v$ for \hcop\ (solid lines) with the histogram of $\delta v$ for HCN overplotted (shaded bars). Dotted lines at $\pm 0.25$ indicate the boundary where $\delta v$ is considered significant (see \S~\ref{s:results}). Right: Comparison of $\delta v$ for \hcop\ and HCN. CepA is labeled because the two tracers have opposite asymmetry. }\label{f:br_hists}
\end{figure}

\section{Comparison with HCN 3-2}\label{s:comparison}
Twenty-four of our 27 sources have also been observed in HCN 3-2 by Wu et al. (2010). Comparing our \hcop\ classification with their HCN classifications, we find that eight of the nine sources we identify as blue asymmetric in \hcop\ (requiring that the line be considered blue by at least two diagnostics) are also blue asymmetric in HCN even though \neff\ (density required to excite a $T_R =1$ K line for gas at $T_{kin}=20$ K, see Evans 1999, Reiter et al. 2011) of the two molecules are different by nearly an order of magnitude (\neff$= 2.3 \times 10^4$ cm$^{-3}$ for \hcop\ 3-2 compared with \neff$ = 1.5 \times 10^5$ cm$^{-3}$ for HCN 3-2 for T$=20$ K). 
Three sources are red asymmetric in both HCN and \hcop. 
The three sources without HCN data are not asymmetric in \hcop\ according to any of our criteria. 
One source, CepA, displays a strong blue asymmetry in HCN ($\delta v = -0.53$) but is red asymmetric in \hcop\ ($\delta v= 0.39$). 
\hcop\ also shows the red asymmetry in the temperature ratio, but the ratio is not significant in HCN. 
The two methods probe slightly different manifestations of the line asymmetry --- the temperature ratio measuring the contrast in the brightness of the blue and red peaks while $\delta v$ measures how far away from the systemic velocity the brighter of the red and blue peaks is maximal. 
That the difference in line asymmetries is most prominently seen in $\delta v$ may reflect the different \neff\ (and, consequently, different inflow velocities) of the two tracers. 
However, given that CepA is red asymmetric by all of our methods, we also consider the possibility that the discrepancy between \hcop\ and HCN may reflect contamination from outflows. 
Multiple outflows have been observed in CepA (Codella et al. 2003, Codella et al. 2005, Torrelles et al. 2011). In addition, the abundance of \hcop\ may be enhanced in outflows (Rawlings et al. 2004), so it is difficult to disentangle the source of the line asymmetry in CepA with just these two tracers. 

Comparing measures of asymmetry between \hcop\ and HCN, we find that $| \delta v |$ is generally larger for the HCN (median $| \delta v_{HCN}|=0.53$, median $| \delta v_{HCO+}|=0.22$, see Figure~\ref{f:br_hists}). 
HCN is excited at an order of magnitude higher density than \hcop, so larger asymmetries may simply reflect a higher inflow velocity because HCN in excited at smaller radii, closer to the center of the clump. 
Chemical enhancement of \hcop\ in outflows may decrease the strength of the \hcop\ asymmetry in general. Half of the sample have CO outflows identified by Wu et al. (2004) and three-fourths of the sample have outflows identified in the literature (e.g. G31.41, Gibb, Wyrowski \& Mundy 2004; G35.20, De Buizer 2006; W75N, Carrasco-Gonz\'{a}lez et al. 2010, Gibb, Davis \& Moore 2007; as well as CepA). 
Outflow motions undoubtedly contribute to the observed \hcop\ line profiles, although disentangling the relative contributions of outflow and inflow motions of multiple cores likely contained within the 30\arcsec\ beam requires multi-dimensional, detailed radiative transfer modeling and high angular resolution observations. The observed excess of blue profiles suggests that inflow motions may dominate the line profile in the eight sources with a blue asymmetry in both lines, given that outflows should produce equal numbers of red and blue profiles.

In addition, different mechanisms may dominate on different scales. Observations of lower \neff\ transitions of \hcop\ (\hcop\ 1-0; Sun \& Gao 2009) reveal the same asymmetry in six of the ten sources common to the two studies 
and four that do not agree. Two sources switch asymmetry between the 1-0 and 3-2 transition while two sources do not show strong asymmetry in one of the two \hcop\ transitions, but are strongly asymmetric in the other. 
Two sources with contradictory \hcop\ profiles are considered strong candidates for collapse by Sun \& Gao (2009), further affirming that multiple transitions probing a range of \neff\ are necessary to assess the likelihood of collapse. 
And while more studies have examined outflows from these sources (e.g. De Buizer 2006, Carrasco-Gonz\'{a}lez et al. 2010), detailed studies of single sources have found compelling evidence for inflow in massive star forming regions (e.g. Keto \& Wood 2006).

Six of the 27 sources also show self-absorption in the HCN isomer HNC J$=3\rightarrow2$ (using data from Reiter et al. 2011). 
The five blue profiles and one red profile are all skewed in the same direction in both HCN and \hcop\ and represent the largest asymmetries in the sample ($ | \delta v_{HCO+} | \sim 0.4$). 
Different spectral resolution of the observations ($0.11$ km s$^{-1}$ for HCN compared with $1.1$ km s$^{-1}$ for HNC) likely prohibits detection of self-absorption in sources with less pronounced asymmetry. 

We might also expect fewer self-absorbed profiles in HNC due the different abundances, and therefore optical depth, of the isomer ($\sim 0.54 - 4.5$; Hirota et al. 1998). 
Even though these two transitions are excited in the same gas and have nearly identical \neff\ 
($1.5 \times 10^5$ cm$^{-3}$ for HCN 3-2 and $1.7 \times 10^5$ cm$^{-3}$ for HNC 3-2), 
they may reflect motions in slightly different columns of material due to differences in optical depth.

\section{Discussion}

Unlike the low mass sources studied by Gregersen et al. (1997) and Mardones et al. (1997), these sources are high mass clumps that likely have multiple cores embedded in the larger cloud. 
Adopting the language of Wu \& Evans (2003), we argue that the eight sources displaying a blue asymmetric profile in both HCN and \hcop\ are candidates for inflow in order to make clear that this is not evidence of the gravitational collapse expected for a single core (Shu 1977). 
Single dish observations in two tracers are not sufficient to confirm inflow, however, the eight sources that show a blue asymmetry in both lines are strong inflow candidates and merit follow-up observations to test this hypothesis. 
Higher resolution maps will be useful for disentangling contributions to the observed line profile as blue asymmetries due to infall should be strongest towards the center of a source (a single core) while blue peaks due to outflows may peak some distance away from the continuum peaks (Mardones et al. 1997). 

Ultimately, identifying the best tracers and candidates for inflow in high-mass regions lays the ground work for higher resolution studies with instruments like ALMA that will be able to resolve individual clumps. 
Hundreds of sources with asymmetric (and possibly self-absorbed) \hcop\ 3-2 line profiles have been uncovered in molecular line follow up of sources discovered in the Bolocam Galactic Plane Survey (Schlingman et al. 2011, in press, BGPS; Aguirre et al. 2011). Of the 1882 sources already observed in \hcop, $\sim11$\% show possible self-absorption. The full sample of $\sim6000$ sources should provide hundreds of inflow candidates for high spatial resolution follow-up.


\section*{Acknowledgments}
We thank Wayne Schlingman and Patrick Fimbres for their assistance with the HHT observations. 
Yancy Shirley is partially supported by NSF grant AST-1008577.




\begin{landscape}

\begin{deluxetable}{lccccccc}
\tablecolumns{7}
\tabletypesize{\scriptsize}
\tablecaption{\hcop\ J=3$\rightarrow$2 Line Profile Classification \label{t:sources}}
\tablewidth{0pt} 
\tablehead{
\colhead{Source}                  &
\colhead{By eye}      &
\colhead{$T_R^*(B) / T_R^*(R)$}      &
\colhead{skew\tablenotemark{a}}               &
\colhead{skew}               &
\colhead{$\delta v$}  &  
\colhead{$\delta v$} & 
\colhead{extent\tablenotemark{b}} \\ 
\colhead{}                  &
\colhead{classification}      &
\colhead{}                  &
\colhead{}               &
\colhead{classification}               &
\colhead{}  &  
\colhead{classification} &
\colhead{\arcsec}
}
\startdata 
121.30+0.66  &	B  &  1.30(.05)  &  -0.26(.07)  &  B  &  -0.54(.07)  &  B  &  44.72  \\ 
123.07-6.31  &	B  &  1.71(.13)  &  -0.44(.10)  &  B  &  -0.20(.03)  &  N  &  22.36  \\ 
W3(OH)  &  B  &  1.41(.06)  &  -0.48(.07)  &  B  &  -0.23(.05)  &  N  &  41.23  \\ 
S231  &  N  &  \nodata  &  0.45(.07)  &  R  &  0.40(.03)  &  R  &  \nodata  \\ 
S252A  &  N  &  \nodata  &  0.20(.09)  &  X  &  0.32(.05)  &  R  &  \nodata  \\ 
RCW142  &  N  &  \nodata  &  -0.26(.04)  &  B  &  -0.10(.03)  &  N  &  \nodata  \\ 
W28A2  &  N  &  \nodata  &  0.24(.04)  &  R  &  0.18(.02)  &  N  &  \nodata  \\ 
M8E  &  R  &  0.94(.05)  &  -0.13(.10)  &  X  &  -0.08(.08)  &  N  &  28.28  \\ 
 8.67-0.36  &  B  &  2.22(.40)  &  -0.57(.08)  &  B  &  -0.44(.05)  &  B  &  30.00  \\
10.60-0.40  &  B  &  2.28(.10)  &  -0.50(.07)  &  B  &  -0.22(.02)  &  N  &  22.36  \\ 
12.89+0.49  &  B  &  1.76(.18)  &  -0.34(.20)  &  X  &  -0.45(.13)  &  B  &  22.36  \\ 
W33A  &  N  &  \nodata  &  0.14(.09)  &  X  &  0.07(.07)  &  N  &  \nodata  \\ 
24.49-0.04  &  N  &  \nodata  &  -0.03(.06)  &  X  &  -0.07(.03)  &  N  &  \nodata  \\ 
W43S  &  N  &  \nodata  &  0.10(.06)  &  X  &  -0.10(.04)  &  N  &  \nodata  \\  
31.41+0.31  &  B  &  1.26(.37)  &  -1.01(.33)  &  B  &  -0.65(.13)  &  B  &  20.00  \\ 
W44  &  B  &  4.73(.35)  &  -0.46(.03)  &  B  &  -0.27(.02)  &  B  &  41.23  \\
40.50+2.54  &  N  &  \nodata  &  -0.06(.14)  &  X  &  -0.01(.07)  &  N  &  \nodata  \\
35.20-0.74  &  R  &  0.71(.04)  &  0.22(.12)  &  X  &  0.60(.05)  &  R  &  31.62  \\ 
59.78+0.06  &  N  &  \nodata  &  -0.34(.05)  &  B  &  -0.09(.06)  &  N  &  \nodata  \\ 
ON1  &  N  &  \nodata  &  0.66(.23)  &  X  &  0.28(.10)  &  R  &  \nodata  \\
ON2S  &  N  &  \nodata  &  0.08(.14)  &  X  &  0.13(.09)  &  N  &  \nodata  \\
W75N  &  R  &  0.70(.01)  &  0.07(.15)  &  X  &  0.35(.06)  &  R  &  31.62  \\
W75OH  &  B  &  1.94(.06)  &  -0.97(.08)  &  B  &  -0.39(.01)  &  B  &  58.31  \\
S140  &  N  &  \nodata  &  -0.10(.08)  &  X  &  -0.14(.06)  &  N  &  \nodata  \\ 
CepA  &  R  &  0.84(.02)  &  0.31(.07)  &  R  &  0.39(.03)  &  R  &  28.28  \\ 
NGC7538-IRS9  &	 N  &  \nodata  &  0.16(.12)  &  X  &  0.06(.07)  &  N  &  \nodata  \\ 
S157  &  N  &  \nodata  &  -0.33(.06)  &  B  &  -0.07(.04)  &  N  &  \nodata  \\
\enddata
\tablenotetext{a}{X = skew is not significant ($\geq 3$ times the standard deviation) using $\pm 1.25$ km s$^{-1}$ from v$_{LSR}$.} 
\tablenotetext{b}{We take the median extent of the asymmetric line profile to account for line asymmetries that persist to different radii in different directions. }
\end{deluxetable}

\end{landscape}

\begin{deluxetable}{lcccccc}
\tablecolumns{5}
\footnotesize
\tablecaption{New H$^{13}$CO$^+$ $J=3\rightarrow 2$ Observed Source Properties \label{t:h13cop32_properties}}
\tablewidth{0pt} 
\tablehead{
\colhead{Source}  &
\colhead{$\alpha$ (J2000.0)}      &
\colhead{$\delta$ (J2000.0)}      &
\colhead{$v_{LSR}$}       & 
\colhead{$I(T_{mb})$}          &
\colhead{$T_{mb}$}       & 
\colhead{$\Delta v$}  \\ 
\colhead{} &
\colhead{} &
\colhead{} &
\colhead{(km/s)} & 
\colhead{(K km/s)} &
\colhead{(K)} & 
\colhead{(km/s)} 
}
\startdata
121.30+0.66  &  00 36 47.5  &  +63 29 01  &  -17.56(0.01)  &  2.49(0.05)  &  0.84(0.01)  &  2.73(0.03)  \\   
123.07-6.31  &  00 52 25.1  &  +56 33 54  &  -30.43(0.05)  &  1.52(0.06)  &  0.38(0.01)  &  3.60(0.11)  \\  
S231  &  05 39 12.9  &  +35 45 54  &  -16.18(0.05)  &  1.65(0.08)  &  0.51(0.02)  &  2.93(0.13)  \\
S252A  &  06 08 35.4  &  +20 39 03  &  9.29(0.04)  &  1.30(0.09)  &  0.53(0.02)  &  2.30(0.12)  \\
24.49-0.04  &  18 36 05.3  &  -07 31 22  &  110.23(0.07)  &  1.57(0.08)  &  0.30(0.01)  &  4.56(0.17)  \\
W75OH  &  20 39 00.9  &  +42 22 50  &  -3.31(0.02)  &  7.77(0.06)  &  1.49(0.01)  &  4.99(0.03)  \\   
S157  &  23 16 04.3  &  +60 01 41  &  -44.23(0.07)  &  0.61(0.07)  &  0.20(0.01)  &  2.51(0.19)  \\  
\enddata
\end{deluxetable}

\end{document}